\documentstyle[aps,multicol,epsf]{revtex}

\begin{document}
\draft

\title{Distribution of the area enclosed by a 2D random walk in a disordered
       medium}

\author{K.V. Samokhin$^*$}

\address{Cavendish Laboratory, University of Cambridge, Madingley Road,
         Cambridge CB3 0HE, UK}
  
\date{\today}

\maketitle

\begin{abstract}
The asymptotic probability distribution for a Brownian particle wandering 
in a 2D plane with random traps to enclose the algebraic area $A$ by time $t$ 
is calculated using the instanton technique. 
\end{abstract}
\bigskip

\begin{multicols}{2}
\narrowtext

It is well known that the properties of random walks are dramatically changed 
by the presence of quenched disorder. In particular, if a walker can be 
irreversibly trapped at some randomly distributed sites, then the asymptotic 
probability to return to the starting point (or the total probability to 
survive until time $t$) is given in $d$ dimensions by $p(t)\sim\exp
(-ct^{d/(d+2)})$ \cite{BV74}, while the mean square displacement decreases 
compared to a pure diffusion: $\langle r^2\rangle\sim t^{2/(d+2)}$ 
\cite{GP82}. On another hand, if a walker is advected by a random force, then 
the total probability is conserved, but the diffusion coefficient and the 
mean square displacement acquire logarithmic corrections \cite{RVF}. Further
examples can be found, for instance, in Ref. \cite{BG90}. Much less, however,
is known about the influence of a quenched disorder on the {\em topological}
properties of random walks, which became a subject of theoretical investigation
very recently \cite{Sam98}. By ``topological'' properties we mean such 
characteristics as the winding number of a Brownian particle, or the linking 
number of a closed polymer, or the algebraic area enclosed by the trajectory
of a random walker. In this Rapid Communication, we concentrate on the latter 
case and calculate the asymptotic distribution of the area swept by a planar 
random walk wandering in the presence of traps.  

The probability distribution for a random walk starting at some point 
${\bf r}'$ at $t=0$ to end at a point ${\bf r}$ after time $t$ obeys the 
diffusion equation
\begin{equation}
\label{diff_eq}
 \frac{\partial P}{\partial t}=D\nabla^2P-U({\bf r})P.
\end{equation}
Here $D$ is the diffusion coefficient and $U({\bf r})=U_0\sum_i\delta({\bf r}-
{\bf R}_i)$ is the random ``potential'', which represents the trapping 
probability per unit time ($U_0>0$). The positions ${\bf R}_i$ of traps are 
distributed uniformly in a plane according to the Poisson law with mean 
density $\rho$. The probability for a random walk of ``length'' $t$ to 
enclose the algebraic area $A$ can be obtained by averaging the corresponding
$\delta$-function constraint over the solutions of Eq. (\ref{diff_eq}) in a 
given distribution of traps:
$$
 {\cal P}(A,t|U)\equiv\left\langle \delta\left(A-\frac{1}{2}\int_0^td\tau\;
  r^2(\tau)\dot\theta(\tau)\right)\right\rangle_{P({\bf r},t;{\bf r}',0|U)},
$$
where $\theta(t)$ is the angle between the radius-vector ${\bf r}(t)$ of the
particle and some fixed direction in the plane. Writing the $\delta$-function 
as an integral over an auxiliary variable $B$, and using the Wiener 
path-integral formula for the solution of Eq. (\ref{diff_eq}), we arrive at 
the following expression:
\begin{equation}
\label{Wiener}
 {\cal P}(A,t|U)=\int_{-\infty}^\infty dp\;e^{iBA}
  \int_{{\bf r}(0)={\bf r}'}^{{\bf r}(t)={\bf r}}{\cal D}{\bf r}(\tau)\,
  e^{-S[{\bf r}(\tau)]},
\end{equation}
where
$$
 S=-\int_0^t d\tau\;\left(\frac{1}{2D} \dot{\bf r}^2(\tau)+U({\bf r}(\tau))+
 \frac{i}{2}Br^2(\tau)\dot\theta(\tau)\right). 
$$
It is easy to see that the path integral on the right-hand side of Eq. 
(\ref{Wiener}) represents the Euclidean Green function 
${\cal G}_B({\bf r},t;{\bf r}',0)$ of a fictitious quantum particle of mass 
$m=(2D)^{-1}$ moving in the random potential
$U({\bf r})$ and in the uniform magnetic field $B$ (we choose the units in 
which $\hbar=e=c=1$). Let us now assume that the trajectory is closed 
(${\bf r}={\bf r}'$) and average the enclosed area distribution over all 
positions of the starting point: $\langle(...)\rangle_{\bf r}=
(1/\Omega)\int d^2r\,(...)$ ($\Omega$ is the system volume). We have:
$$
 {\cal P}(A,t|U)=\int_{-\infty}^\infty dB\;e^{iBA}Z(t,B|U),
$$  
where $Z(t,B)$ is the partition function at inverse ``temperature'' $1/T=t$. 
Averaging now over the positions of traps, we finally obtain: 
\begin{equation}
\label{gen_P}
 {\cal P}(A,t)=\int_{-\infty}^\infty dB\int_0^\infty dE\;e^{iBA}e^{-Et}N(E,B),
\end{equation}
where $N(E,B)$ is the average density of states of a quantum particle 
described by the Hamiltonian
\begin{equation}
\label{Hamilt}
 H=D(-i\nabla-{\bf A}({\bf r}))^2+U({\bf r}).
\end{equation}
We choose the cylindrical gauge for the vector potential: $A_\theta=Br/2$. 

In an ideal case (i.e. in the absence of traps), the eigenvalues of the 
Hamiltonian (\ref{Hamilt}) are the Landau levels $E_n=\omega_{\rm c}(n+1/2)$,
where $\omega_{\rm c}=2DB$ is the cyclotron frequency. The density of states
is then given by the set of equidistant $\delta$-function peaks, and the 
partition function can be easily evaluated, resulting in the following 
expression for the enclosed area distribution:  
\begin{equation}
\label{Levy}
 {\cal P}(A,t)\sim\frac{1}{\cosh^2x},\qquad x\sim\frac{A}{Dt}.
\end{equation}
This result was first obtained by Levy quite a while ago \cite{Levy48}.
The form of the dimensionless scaling variable $x$ is quite natural, since 
the only characteristic scale with dimensionality of area in a clean system 
is the mean square displacement $\langle r^2\rangle\sim Dt$.

In order to calculate the density of states in the presence of disorder, let
us first estimate what characteristic scales of energy and magnetic field
determine the asymptotic behavior of ${\cal P}(A,t)$. We are interested in 
calculation of the probability distribution at large but fixed $t$ and 
$A\to\infty$. As seen from (\ref{gen_P}), this limit corresponds to
$E\to 0$, $B\to 0$, and $\omega_{\rm c}\ll E$. Therefore, it looks natural to 
start with the case of $B=0$ and make sure that we are able to treat the 
magnetic field as a small perturbation in the relevant range of parameters.

At $B=0$, the low-energy behavior of the density of states is determined by
the rare fluctuations of the concentration of impurities creating the large
areas free of traps which are able to sustain the eigenstates with $E\to 0$. 
The asymptotic expression is given by $N(E,B=0)
\sim\exp(-{\rm const}\,\rho D/E)$ at $E\ll E_0$, where $E_0=\rho U_0$ is the 
mean value of the random potential \cite{Lif65}. Quantitatively, such 
exponentially small ``tails'' are determined by the contributions of 
instantons, i.e. spatially localized solutions of the saddle-point equations 
of the effective field theory. The typical size of a clean area, or the 
instanton diameter, grows in the time representation as $l_{\rm inst}\sim
(Dt/\rho)^{1/4}$ \cite{GP82}. If an external magnetic 
field is imposed on the instanton, then its effect on the energy spectrum 
can be calculated perturbatively as long as the magnetic length 
$l_B=\sqrt{1/B}$ considerably exceeds the instanton dimension $l_{\rm inst}$, 
which is the case if $A\gg(Dt/\rho)^{1/2}$. It is this condition that 
determines the limits of applicability of our theory. Note also that if we 
were interested in calculation of the intermediate asymptotics of 
${\cal P}(A,t)$ at moderate $A$, then we could use the exact expressions 
for the density of states at $E-\omega_{\rm c}/2\ll\omega_{\rm c}$ 
\cite{BGI84} (since the random potential is positive everywhere, the density 
of states vanishes at $E<\omega_{\rm c}/2$).

Let us now calculate the effect of an external magnetic field on instantons
explicitly. The density of states can be expressed in terms of the retarded
Green function of the Schr\"odinger equation with the Hamiltonian 
(\ref{Hamilt}): $N(E,B)=-(\pi\Omega)^{-1}\int d^Dr\;{\rm Im}\,\langle 
G^R({\bf r},{\bf r};E,B)\rangle_U$, where $G^R({\bf r},{\bf r}';E,B)=
\langle {\bf r}|(E-H+i0)^{-1}|{\bf r}'\rangle$. The Green function can, in 
turn, be calculated by standard means of the quantum field theory. In order to 
carry out the averaging over the positions of traps, we resort to the 
supersymmetry approach, in which the cancellation of denominators is achieved
by doubling the degrees of freedom and introducing the commuting and 
anticommuting fields on equal footing (see, for instance, Ref. \cite{Efet97}).
Before disorder averaging, the retarded Green function can be written as the
following functional integral:
\begin{eqnarray}
\label{func_int}
 G^R({\bf r},{\bf r}';E,B)&=&-i\lim\limits_{\eta\to +0}\int{\cal D}^2
  \Phi({\bf r})\;\varphi({\bf r})\bar\varphi({\bf r}')\nonumber\\
  &\times&\exp\left(i\int d^2r\;\bar\Phi(E-H+i\eta)\Phi\right),
\end{eqnarray}
where the two-component superfields  
$$
 \Phi({\bf r})={\varphi({\bf r})\choose\psi({\bf r})},\qquad 
  \bar\Phi({\bf r})=(\varphi^*({\bf r}),\bar\psi({\bf r}))
$$
are composed of a Bose field $\varphi$ and a Grassmanian field $\psi$, and
${\cal D}^2\Phi=(1/\pi){\cal D}({\rm Re}\,\varphi){\cal D}({\rm Im}\,\varphi)
{\cal D}\bar\psi{\cal D}\psi$. The probability of having $N$ impurities 
located at the points ${\bf R}_1,...,{\bf R}_N$ in the area $\Omega$ in the 
plane is given by the Poisson law:
$$
 P_N({\bf R}_1,...,{\bf R}_N)=\frac{e^{-\rho\Omega}}{N!}(\rho\Omega)^N.
$$
Using this expression to average (\ref{func_int}) over the positions of traps,
we end up with the following effective action:
\begin{eqnarray}
\label{im_action}
 iS[\Phi]=\int d^Dr\;\Bigl\{&& i\bar\Phi\left(E-D(-i\nabla-{\bf A})^2+i\eta
 \right)\Phi \nonumber \\
 && -\rho\left(1-e^{-iU_0\bar\Phi\Phi}\right)\Bigr\}. 
\end{eqnarray}

An important property of the action (\ref{im_action}) is that it is invariant 
under the global supersymmetry transformations, mixing the boson and fermion 
sectors:
\begin{equation}
\label{s_symm}
 \Phi\to\tilde\Phi=T\Phi,
\end{equation}
where $T$ is a $2\times 2$ unitary supermatrix \cite{Efet97}. Due to this 
symmetry, the problem of finding the saddle points of (\ref{im_action}) can 
be considerably simplified. Indeed, since the saddle point manifold is 
invariant under the transformations (\ref{s_symm}), we are able to seek the 
instanton solution in the following form:
\begin{equation}
\label{gen_sp}
 \Phi_{\rm inst}({\bf r})={\varphi_{\rm inst}({\bf r})\choose 0},
\end{equation} 
where $\varphi_{\rm inst}({\bf r})=\varphi_{\rm inst}(r)$ is a cylindrically 
symmetric spatially localized function. If to write the boson fields as 
$\varphi=\varphi_1+i\varphi_2$, $\varphi^*=\varphi_1-i\varphi_2$, where 
$\varphi_{1,2}$ are real on the initial functional integration contour, then 
the imaginary part of the Green function is determined by a non-trivial 
saddle point of the action in the complex plane of $\varphi_{1,2}$ 
\cite{Lang67,CC77}, and
\begin{equation}
\label{DoS_inst}
 N(E,B)\sim e^{-S_{\rm inst}(E,B)}
\end{equation}
with the exponential accuracy. The fermion sector can be neglected as long as 
we are not interested in calculation of the pre-exponential factor in 
(\ref{DoS_inst}).  

In order to make Eq. (\ref{im_action}) in the boson sector real, we rotate 
the integration contour: $\varphi_{1,2}\to e^{-i\pi/4}\varphi_{1,2}$.
As result of this, the exponent on the right-hand side of Eq. (\ref{func_int}) 
changes: $iS[\Phi]\to -S[\Phi]$, where the action $S$ is
\begin{eqnarray}
\label{S_B}
 S=\int d^2r\;\Bigl\{&&\varphi_i\left(-E+D(-i\nabla-{\bf A})^2\right)
 \varphi_i \nonumber\\
 &&+\rho\left(1-e^{-U_0\varphi_i^2}\right)\Bigr\}
\end{eqnarray}
($i=1,2$). Introducing the dimensionless variables:
$$
 r=\xi x,\qquad \varphi_{\rm inst}(r)=\sqrt{U_0} f(x),
$$ 
where $\xi=\sqrt{D/E}$, we obtain, from Eq. (\ref{S_B}), a non-linear 
differential equation for the saddle-point solution $f$:
\begin{equation}
\label{gen_eq}
 -\frac{1}{x}\frac{d}{dx}\left(x\frac{d}{dx}\right)f+\beta^2x^2f+
  \alpha^2e^{-f^2}f=f,
\end{equation}
where $\alpha=\sqrt{E_0/E}\gg 1$ and $\beta=\omega_{\rm c}/4E\ll 1$. Similar
equations without magnetic field were obtained in Refs. \cite{FL75,Lub84}, 
using different techniques. 

Due to complexity of the equation (\ref{gen_eq}), we are able to find only an 
approximate solution. To this end, we replace the ``potential'' 
$V(f)=\alpha^2e^{-f^2}$ in Eq. (\ref{gen_eq}) by a piecewise constant 
potential:
\begin{equation}
 V_{\rm eff}(f)=\left\{ \begin{array}{ll}
  \alpha^2\ &, \mbox{ at }f<1, \\
  0\ &, \mbox{ at }f>1.
  \end{array} \right.  
\end{equation}
Then, Eq. (\ref{gen_eq}) becomes effectively linear and reduces to a couple of 
the Schr\"odinger equations, whose solutions $f_{1,2}(x)$ satisfy the 
following conditions: 
\begin{equation}
\label{bcs} 
 f_1(x_1)=f_2(x_1)=1,\ f_1'(x_1-0)=f_2'(x_1+0),
\end{equation}
where $x_1(\alpha,\beta)$ is the position of the discontinuity in the 
effective potential, which is to be determined self-consistently.  
Going back to the dimensional variables, it is easy to see from (\ref{S_B}) and
(\ref{gen_eq}) that the instanton action is proportional to the area of the
effective potential well:
\begin{equation}
\label{S_inst}
 S_{\rm inst}=\pi\rho\xi^2x_1^2.
\end{equation} 

Let us start with the case of $B=0$, i.e. $\beta=0$. The solution of the 
linearized saddle-point equations can be written in form
\begin{equation}
\label{zero_sol}
 f(x)=\left\{ \begin{array}{ll}
   f_1(x)=C_1J_0(x),&\ \mbox{at }0<x<x_1,\\
   f_2(x)=C_2K_0(\alpha x),&\ \mbox{at }x_1<x,
  \end{array} \right. 
\end{equation}
where $J_0(x)$ and $K_0(x)$ are the Bessel functions of real and imaginary 
arguments, respectively. Substituting this solution in the matching conditions
(\ref{bcs}), we obtain the equation
\begin{equation}
 \frac{J_1(x)}{J_0(x)}=\alpha\frac{K_1(\alpha x)}{K_0(\alpha x)}.
\end{equation}
In the limit $\alpha\to\infty$, assuming that $x_1\sim 1$ and using the 
asymptotic expansions of the Bessel functions \cite{AS65}, we obtain 
$J_0(x_1)=0$, i.e. $x_1=a$, where $a\approx 2.405$ is the first zero of the 
function $J_0(x)$. After substitution in (\ref{S_inst}) and (\ref{DoS_inst}), 
the Lifshitz result $N(E)\sim\exp(-{\rm const}\,\rho D/E)$ is recovered.

In principle, one can find the exact solutions of the Schr\"odinger equations 
inside and outside the potential well at $B\neq 0$ (they are expressed in 
terms of the confluent hypergeometric functions), match them at the point 
$x=x_1$ and finally end up with a transcendental equation for 
$x_1(\alpha,\beta)$, which can be solved at $\beta\to 0$.  
However, we prefer not to follow this procedure here, because the same results
can be obtained using much more physically apparent reasoning in the spirit 
of the Lifshitz's original derivation \cite{Lif65}. First, we note that, 
in the limit $\alpha\to\infty$, the instanton solution satisfies the 
Schr\"odinger equation for a particle confined in the potential well 
with infinitely high walls in a magnetic field. 
The ground state energy $\epsilon_0$ is equal to 
unity (in the units of $E$). In the absence of magnetic field this condition 
fixes the radius of the well at $x_1(0)=a$. At $\beta\neq 0$, the lowest order 
perturbative correction to the ground state energy is
\begin{equation}
 \delta\epsilon_0=\frac{\displaystyle\beta^2\int_0^a dx\,x^3f^2(x)}{
  \displaystyle\int_0^a dx\,xf^2(x)},
\end{equation}
where $f(x)\sim J_0(x)$ is the unperturbed ground state wave function 
(\ref{zero_sol}). Calculating the integrals with the Bessel functions, we 
obtain: $\delta\epsilon_0=c\beta^2$, where $c\simeq 1.261$. To keep the ground
state energy fixed, this correction should be compensated by the corresponding
increase in the radius of the potential well: $x_1^2(\beta)=x_1^2(0)(1+\delta
\epsilon_0)$. Substituting this in (\ref{S_inst}), we finally obtain:
\begin{equation}
\label{result_action} 
 S_{\rm inst}(E,B)\simeq\frac{\pi\rho Da^2}{E}\left(1+\frac{cD^2}{4E^2}B^2
 \right).
\end{equation} 
This expression is valid at $E\to 0$, $B\to 0$, $\omega_{\rm c}/E\to 0$. 

The asymptotic probability distribution of the enclosed area can now be
obtained from (\ref{gen_P}), (\ref{DoS_inst}) and (\ref{result_action}) by 
calculating the integrals by the steepest descent method:
\begin{eqnarray}
\label{result}
 {\cal P}(A,t)&\sim&\int_{-\infty}^\infty dB\int_0^\infty dE\;e^{iBA}e^{-Et}
  e^{-S_{\rm inst}(E,B)}\nonumber\\
 &\sim& \exp(-\sqrt{\pi\rho Da^2t})\exp\left\{-\frac{a\sqrt{\pi}}{c}
  \frac{\sqrt{\rho}A^2}{(Dt)^{3/2}}\right\}.            
\end{eqnarray}
The first exponential on the right-hand side is nothing but the asymptotic 
``tail'' of the total probability $p(t)=\int dA\,{\cal P}(A,t)$ for a random
walker not to be trapped \cite{BV74}. The second exponential thus represents 
the conditional probability to enclose the area $A$, provided a walker has 
survived until time $t$.

We see that the asymptotic behavior of ${\cal P}(A,t)$ is drastically changed
by the presence of disorder, compared to the Levy's result (\ref{Levy}). The
distribution becomes Gaussian ${\cal P}(A,t)\sim\exp(-x^2)$, with the scaling 
variable
\begin{equation}
 x\sim\frac{A}{Dt}\,(\rho Dt)^{1/4},   
\end{equation}
so that the standard deviation now grows slower than in the clean case:
$\langle A^2\rangle^{1/2}\sim t^{3/4}$ (the mean value $\langle A\rangle$ 
is, of course, zero). Such a different form of the scaling variable can be 
related to the presence of an extra length scale $r_{\rho}\sim\sqrt{1/\rho}$ 
in the system, which depends on the concentration of traps (nothing depends 
on the absolute value $U_0$ of the random potential). Note also that a 
similar Gaussian distribution was obtained in Ref. \cite{DC92} for a 2D random
walk in a box of a finite size $L$. In this case, the fictitious magnetic 
field was also treated perturbatively, giving rise to $\langle 
A^2\rangle^{1/2}\sim L(Dt)^{1/2}$. Qualitatively, such a difference between 
the two systems is due to the fact that in our case the size of the effective 
potential well is not constant, but grows with time as $L(t)=l_{\rm inst}(t)
\sim t^{1/4}$. 

In conclusion, we studied the asymptotic probability distribution of the 
algebraic area enclosed by a planar random walk in the presence of immobile 
random traps. It is shown that this probability is directly related to ``the
Lifshitz tail'' in the density of states of a quantum particle in a 
Poisson disorder and uniform magnetic field. In contrast to the case of an 
ideal random walk, the enclosed area distribution turns out to be Gaussian 
with the standard deviation growing as $t^{3/4}$.  
\\  

This work was financially supported by the Engineering and Physical Sciences
Research Council (UK).

\end{multicols}

\end{document}